# Statistical methods for characterizing transfusion-related changes in regional oxygenation using near-infrared spectroscopy (NIRS) in preterm infants

Ying Guo,[1]  Yikai Wang,[1] Terri Marin,[2] Easley Kirk,[1] Ravi M Patel[3] and Cassandra D Josephson[3]

## Abstract

Near infrared spectroscopy (NIRS) is an imaging-based diagnostic tool that provides non-invasive and continuous evaluation of regional tissue oxygenation in real-time. In recent years, NIRS has shown promise as a useful monitoring technology to help detect relative tissue ischemia that could lead to significant morbidity and mortality in preterm infants. However, some issues inherent in NIRS technology use on neonates, such as wide fluctuation in signals, signal dropout and low limit of detection of the device, pose challenges that may obscure reliable interpretation of the NIRS measurements using current methods of analysis. In this paper, we propose new nonparametric statistical methods to analyze mesenteric $rSO_2$ (regional oxygenation) produced by NIRS to evaluate oxygenation in intestinal tissues and investigate oxygenation response to red blood cell transfusion (RBC) in preterm infants. Specifically, we present a mean area under the curve (MAUC) measure and a slope measure to capture the mean $rSO_2$ level and temporal trajectory of $rSO_2$, respectively. We develop estimation methods for the measures based on multiple imputation and spline smoothing and further propose novel nonparametric testing procedures to detect RBC-related changes in mesenteric oxygenation in preterm infants. Through simulation studies, we show that the proposed methods demonstrate improved accuracy in characterizing the mean level and changing pattern of mesenteric $rSO_2$ and also increased statistical power in detecting RBC-related changes, as compared with standard approaches. We apply our methods to a NIRS study in preterm infants receiving RBC transfusion from Emory Univerity to evaluate the pre- and post-transfusion mesenteric oxygenation in preterm infants.

## 1 Introduction

Near infrared spectroscopy (NIRS) technology is a diagnostic tool to non-invasively and continually evaluate regional tissue oxygenation in real-time.[1] This type of monitoring has particular clinical significance as relative tissue ischemia is captured by NIRS producing decreased $rSO_2$ (regional oxygenation) measurements.[1] The application of NIRS in very low birth weight (VLBW) infants has great appeal, especially due to the non-invasive application and continuous monitoring, as these infants are vulnerable to diminished oxygen delivery to many vital tissues including cerebral, mesenteric, and renal.[1–5] The consequences related to inadequate oxygenation of these organs causes significant morbidity and mortality in preterm infants. One such morbidity is necrotizing enterocolitis (NEC), which is an intestinal inflammatory disease with a multifactorial etiology.[6,7] Factors contributing to NEC include abnormal intestinal microbiota and ischemia, possibly from inadequate

[1]Department of Biostatistics & Bioinformatics, Emory University, Atlanta, GA, USA
[2]Department of Physiology and Technology, Augusta State University, Augusta, GA, USA
[3]School of Medicine, Emory University, Atlanta, GA, USA

**Corresponding author:**
Ying Guo, Department of Biostatistics & Bioinformatics, Emory University, Atlanta, GA, USA.
Email: yguo2@emory.edu

oxygen delivery from anemia or abnormal vascular regulation.[8,9] Given the high-case fatality rate and limited treatment options once an infant develops NEC, new tools and approaches are greatly needed to identify infants who are most likely to develop NEC or have worse outcomes once they have the disease. Towards this end, several studies have started utilizing NIRS in assessing mesenteric oxygenation in the setting of VLBW preterm infants.[10,11] Compared with alternative tools such as color Doppler and ultrasonography which cannot be performed continuously to follow evaluation of disease,[12] NIRS provides a significant advantage for monitoring real-time changes in oxygen extraction within tissue beds, hence allowing direct and continuous comparison of how much oxygen is being delivered vs. consumed in different (or regional) organ tissue beds.[2] Recent studies have suggested a temporal relationship between red blood cell (RBC) transfusion and NEC onset,[13] but a specific mechanism has yet to be determined. Some reports suggest enteral feeding continuation during RBC transfusion decreases mesenteric oxygenation.[14] Moreover, recent studies have demonstrated that the severity of anemia (hemoglobin per week of $\leq 8\,g/dL$), not RBC transfusion exposure, is associated with NEC.[9] Many prospective observational investigations have started utilizing NIRS to examine these aforementioned relationships.

Although NIRS technology has shown great promises as an effective and non-invasive new tool for monitoring VLBW infants and potentially for early detection of NEC, there is no systematic or consistent methodological approach to analyze the NIRS data output. Moreover, NIRS technology use on neonates poses several challenges that may obscure reliable statistical interpretation using current methods of analysis. For example, raw $rSO_2$ measurements are often highly variable which may mask the underlying temporal changing pattern in oxygen extraction. Second, currently used NIRS devices generally produce an $rSO_2$ measurement range of 15% to 95% as any measurement beyond this scale becomes nonspecific and unreliable. Because of this design, signal "dropout" may occur in severely compromised tissue beds due to the 15% low detection cut-off, which potentially leads to biased results in the subsequent analysis. Furthermore, as with all transcutaneous monitor devices, loose sensor adherence may cause periods of artifact, especially on tenuous skin surfaces of VLBW infants. These issues may lead to imprecise results using conventional analysis of intestinal $rSO_2$, which exhibit greater variability than other tissues beds such as cerebral and renal.[15]

In this paper, we develop a nonparametric statistical framework to characterize infants' mesenteric oxygen saturation and its changes as they relate to RBC transfusion. We propose to model the observed $rSO_2$ time series with a nonparametric smooth function, estimated via penalized regression splines,[16] to capture the underlying biological process of mesenteric oxygen saturation. We present two new statistical measures for accurately characterizing important features in $rSO_2$ series. The first measure is a mean area under the curve (MAUC) statistic that reflects the average $rSO_2$ levels during a specific time period. The second is a slope measure that reflects the rate and direction of temporal change in $rSO_2$ trajectory during the pre- and post-transfusion period. These two measures capture different aspects of temporal dynamics of the regional oxygenation. We present robust estimation approaches and nonparametric tests for the MAUC and slope measure. The proposed inference procedures can adjust for the detection limit problem in the NIRS device, improve accuracy in characterizing the mean level and trajectory of $rSO_2$ and increase statistical power in detecting RBC-related changes in infants' mesenteric oxygen saturation.

Our proposed methods address some of the challenging issues in NIRS $rSO_2$ data analysis, result interpretation, and clinical relevance of findings. Through simulation studies and application to a real-world NIRS data example, we illustrate the advantages of the proposed methods over existing approaches. We have developed an R package *NIRStat* for implementing the methods in the paper and made it available on CRAN.

## 2 Motivating data example

The motivating data were from a previous prospective observational study investigating the effect of RBC transfusion on mesenteric oxygenation measured by NIRS conducted at a level III neonatal intensive care unit (NICU) in Emory University Hospital Midtown at Atlanta, Georgia.[17] The decision to transfuse was determined by the attending neonatologist independent of this study. Infants with congenital anomalies, intraventricular hemorrhage Grade III or greater, hemodynamically significant patent ductus arteriosus, requiring vasopressor support or previous NEC were excluded from this study. Nineteen VLBW infants were enrolled in the study and mesenteric $rSO_2$ NIRS measurements were obtained 30 min prior to each RBC transfusion, during transfusion, and up to 48 h following transfusion completion. Among the 19 enrolled infants, five were excluded from the analysis because they either developed NEC during transfusion or received split transfusions, i.e. two half doses separated by 12 h. For our analysis, we included data collected on the remaining 14 infants who received RBC transfusions. Per study protocol, all transfusions throughout their hospitalization were monitored with mesenteric NIRS and the transfusion approach was guided per the clinical team. Of the 14 infants, four received two transfusions during their hospitalization producing 18 NIRS data sets for analysis. The mean (SD) birth gestational age was 27.6 (2.21) weeks and birthweight was 1039 (256.3) g. Of the 14 infants, 10 were male, 11 were black, and all were appropriate for gestational age.

Mesenteric $rSO_2$ values were measured using an FDA-approved NIRS somatic oximeter (INVOS 5100C; Covidien, Boulder, CO, USA). The INVOS 5100C (Covidien) is a frequently used NIRS device for neonatal clinical research because the $rSO_2$ values produced are 75% venous weighted with 25% arterial and capillary measured, which provides a highly sensitive evaluation of actual tissue oxygenation. NIRS produces an $rSO_2$ value that reflects total oxygen bound to hemoglobin. The measured value is the result of the amount of oxygen delivered minus oxygen consumed at the tissue level, reflecting tissue oxygenation. Data were recorded every 30 s in real-time before, during, and up to 48 h after each RBC transfusion event. NIRS sensor probes were placed on the infant mid-abdomen, below the umbilicus for the mesenteric readings.

## 3 Methods

In this section, we present two statistical measures for characterizing the features in the $rSO_2$ measurements during pre-transfusion and post-transfusion time-period. These two measures capture different aspects in $rSO_2$ temporal pattern.

### 3.1 Mean area under the curve

To set notation, let $Y(t_k)$ denote the $rSO_2$ value measured at time $t_k\{k = 1,\ldots,K\}$ for a subject and $\mathbf{Y} = \{Y(t_k), k = 1,\ldots,K\}$, we assume the following nonparametric model for the observed $rSO_2$ series

$$Y(t_k) = \mu(t_k) + \varepsilon(t_k) \qquad (1)$$

where $\mu(t)$ denotes a smooth function representing the oxygen saturation biological process underlying the observed $rSO_2$ series and $\varepsilon(t)$ is a zero-mean random process representing the variability of the observed data around the underlying smooth biological function. To capture the average oxygen saturation level, we propose a MAUC statistic based on the smooth function $\mu(t)$. Specifically, MAUC of $\mu(t)$ within a time interval (a,b) is defined as follows

$$\text{MAUC}(a,b) = \frac{\int_a^b \mu(t)\mathrm{d}t}{b - a} \qquad (2)$$

The MAUC essentially represents the average mesenteric oxygen saturation level during a given time interval. This is the most commonly used information from NIRS $rSO_2$ series for monitoring infants.

To estimate MAUC, we first obtain a smoothing spline estimate $\hat{\mu}(t)$ based on penalized regression splines[18] where the smoothing parameter is chosen based on Generalized Cross Validation (GCV) criterion.[19] The estimate of MAUC is then obtained from $\hat{\mu}(t)$. One issue in MAUC estimation is that the $rSO_2$ measurements acquired by current NIRS devices have a low detection cut-off at 15% which can potentially cause biased estimate of the average $rSO_2$ level. To address this issue, we consider a multiple imputation approach which has been shown to be an effective strategy to help reduce bias due to limit of detection problem.[20,21]

In the following, we present the algorithm for estimating the MAUC within an interval (a,b) based on a set of observed $rSO_2$ $\mathbf{Y} = \{Y(t_k), k = 1,\ldots,K\}$.

---

**Algorithm 1.** Estimation of MAUC

---

*Step 1*: For the $m$th imputation ($m = 1,\ldots, M$), for an observed $rSO_2$ at the detection limit in the observed $rSO_2$ series $\mathbf{Y}$, i.e. $Y(t_j) = 15\%$, we generate a random value $Y_m^*(t_j)$ from the uniform distribution $U(0,15\%)$ which will be used as the $m$th imputed value for $Y(t_j)$. Repeat the imputation for all the observations at the detection limit in $\mathbf{Y}$. The new set of observations with the imputed values is then denoted as $\mathbf{Y}_m^*$.

*Step 2*: Using the penalized regression spline method, obtain a smoothing spline estimate $\hat{\mu}_m(t)$ based on the imputed data $\mathbf{Y}_m^*$ and then derive the estimate $\text{MAUC}_m^*(a, b)$ based on $\hat{\mu}_m(t)$ following equation (2).

*Step 3*: Repeating Steps 1 and 2 for $M$ times, the imputation-based estimator of MAUC is derived as follows

$$\mathrm{MAUC}^*(a, b) = \frac{1}{M}\sum_{m=1}^{M}\mathrm{MAUC}_m^*(a, b)$$

We also present the following variance estimator for MAUC*(a,b) proposed in Algorithm 1

$$\frac{1}{M}\sum_{m=1}^{M}\mathrm{W}_m^{\mathrm{AUC}} + \left(1 + \frac{1}{M}\right)B_{\mathrm{AUC}}$$

where $B_{\mathrm{AUC}}$ is the between-imputation sample variance of $\mathrm{MAUC}_m^*(a, b)$ (m = 1,...,M) and $\mathrm{W}_m^{\mathrm{AUC}}$ is the variance of $\mathrm{MAUC}_m^*(a, b)$ within its specific imputed dataset which can be estimated by the block bootstrap approach which takes into account the temporal correlations in the $rSO_2$ series.

## 3.2 A nonparametric test for the difference in MAUC between pre- and post-transfusion period

A main goal in the NIRS study is to examine transfusion-related changes in $rSO_2$. To this end, we present a nonparametric permutation test to examine the difference in MAUC between the pre- and post-transfusion period. For a given infant, we denote the pre-transfusion $rSO_2$ time series as $\mathbf{Y}_{\mathrm{pre}} = \{Y_{\mathrm{pre}}(t_k), k = 1,...,K_{\mathrm{pre}}\}$ where $t_1$ is set as 0 and $K_{\mathrm{pre}}$ is the total number of $rSO_2$ measurements recorded during pre-transfusion period for the infant and $\{t_k\}$ is the series of time points when $rSO_2$ values were recorded. We denote the post-transfusion $rSO_2$ time series as $\mathbf{Y}_{\mathrm{post}} = \{Y_{\mathrm{post}}(t_k), k = K_{\mathrm{pre}} + 1,..., K_{\mathrm{pre}} + K_{\mathrm{post}}\}$ where $K_{\mathrm{post}}$ is the total number of $rSO_2$ measurements recorded during post-transfusion period for the infant. Using the method in Section 3.1, we can estimate the $MAUC_{pre}^*$ and $MAUC_{post}^*$ based on the observed $\mathbf{Y}_{\mathrm{pre}}$ and $\mathbf{Y}_{\mathrm{post}}$, respectively. We then evaluate the pre- to post-transfusion changes in the MAUC as

$$\Delta_{MAUC} = MAUC_{post}^* - MAUC_{pre}^*$$

We consider the following permutation test to examine whether there is significant change between the MAUC from the pre- and post-transfusion periods. In the proposed test, we permute the temporal orders in the observed $rSO_2$ series collected within each subject during the pre- and post-transfusion period to generate a null distribution of the test statistics under the null hypothesis that there are no differences in MAUC between pre-transfusion and post-transfusion periods. As mentioned, an issue with the observed $rSO_2$ is the presence of low detection limit. To appropriately account for this matter, we present a multiple imputation (MI) nested permutation test procedure

**Algorithm 2.** a MI nested permutation test

(1) For $g = 1,...G$, (The permutation procedure)

Let $\left\{k_j^{(g)}\right\}_{j=1}^{K_{pre}+K_{post}}$ be the $g$ th permutation of index$\{1,..., K_{pre} + K_{post}\}$, and $\{t_{k_j^{(g)}}\}$ are the corresponding permuted time points, let $\mathbf{Y}_{pre}^{(g)} = \left\{Y_{pre}(t_{k_j^{(g)}})\right\}_{j=1}^{K_{pre}}$ and $\mathbf{Y}_{post}^{(g)} = \left\{Y_{post}(t_{k_j^{(g)}})\right\}_{j=K_{pre}+1}^{K_{pre}+K_{post}}$ denote the pre-transfusion and post-transfusion $rSO_2$ series based on the $g$ th permutation and we further denote $\mathbf{Y}^{(g)} = \{\mathbf{Y}_{pre}^{(g)}, \mathbf{Y}_{post}^{(g)}\}$.

(2) For $m = 1,...,M$, (The multiple imputation procedure)
   (a) For each observed $rSO_2$ at the detection limit 15% in $\mathbf{Y}^{(g)}$, generate a random value from the uniform distribution U(0,15%) and use it as the $m$th imputed value. Repeat the imputation for all the observations at the detection limit in $\mathbf{Y}^{(g)}$. The new set of observations with the imputed values is then denoted as $\mathbf{Y}_m^{(g)*} = \{\mathbf{Y}_{m, pre}^{(g)*}, \mathbf{Y}_{m, post}^{(g)*}\}$.

(b) Obtain an estimate of the $MAUC_{m,pre}^{(g)*}$ and $MAUC_{m,post}^{(g)*}$ for the pre- and post-transfusion periods based on $Y_{m,pre}^{(g)*}$, $Y_{m,post}^{(g)*}$, respectively, and then evaluate the pre- to post-transfusion changes in the MAUC as

$$\Delta_{m,MAUC}^{(g)} = MAUC_{m,post}^{(g)*} - MAUC_{m,pre}^{(g)*}$$

After repeating step a. and b. for $m = 1, \ldots, M$, the MAUC difference estimate for the $g$ th permuted sample is derived as

$$\Delta_{MAUC}^{(g)} = \frac{1}{M} \sum_{m=1}^{M} \Delta_{m,MAUC}^{(g)}$$

(3) After repeating steps (1) and (2) for $g = 1, \ldots, G$, we derive the permutation test $p$-value for $\Delta_{MAUC}$ based on the empirical distribution of $\{\Delta_{MAUC}^{(g)}\}_{g=1}^{G}$ as follows

$$p_{MAUC}^{perm} = \frac{1}{G} \sum_{g=1}^{G} I\left(\left|\Delta_{MAUC}^{(g)}\right| > |\Delta_{MAUC}|\right)$$

## 3.3 The slope of the rSO$_2$ trajectory

The MAUC measure proposed Section 3.1 reflects the average level of rSO$_2$ during a period of time. When monitoring an infant's condition with the NIRS technology, another important feature is how a subject's oxygen level changes across time. The temporal trajectory of the rSO$_2$ measurements can provide information on whether the oxygen level remains stable, is experiencing a fast decrease indicating a quick deterioration in the intestinal condition, or shows significant increase after the transfusion. The temporal pattern is of interest for researchers because it potentially provides a sensitive measure to capture modifications in the status of a baby, where a positive slope with an increase in rSO$_2$ indicates potential improvement in the health status of a baby and a significant negative slope with decrease in rSO$_2$, indicating the condition of the baby is deteriorating.

To capture the trajectory in rSO$_2$ levels, the naïve method is to fit linear models to the observed rSO$_2$ measurements $\mathbf{Y}_{pre}$ and $\mathbf{Y}_{post}$, respectively and then compare the difference in the estimated slopes during pre- and post-transfusion periods. Though straightforward to apply, this naïve approach has several limitations. It does not take into account the NIRS device's low detection cut-off. The linear model is fitted based on raw rSO$_2$ measurements which are highly noisy. This can potentially lead to inaccuracy and high variability in the estimated slopes, which in turn decreases the statistical power in detecting the pre- to post-transfusion changes in the trajectories.

To more reliably capture the trajectory, we propose to model the observed rSO$_2$ series using the nonparametric model in equation (1) and then fit a linear model to the smooth function $\mu(t)$. The slope parameter in the linear model represents the temporal change rate in the underlying biological process. To take into account the detection limit at 15%, we apply the multiple imputation approach. The following algorithm summarizes the steps for estimating the slope parameter based on the observed rSO$_2$ series $\mathbf{Y} = \{Y(t_k), k = 1, \ldots, K\}$.

**Algorithm 3.** Estimation of the slope measure

Step 1: For the $m$th imputation ($m = 1, \ldots, M$), for an observed rSO$_2$ at the detection limit in the observed rSO$_2$ series $\mathbf{Y}$, i.e. $Y(t_j) = 15\%$, we generate a random value $Y_m^*(t_j)$ from the uniform distribution $U(0, 15\%)$ which will be used as the $m$th imputed value for $Y(t_j)$. Repeat the imputation for all the observations at the detection limit in $\mathbf{Y}$. The new set of observations with the imputed values is denoted as $\mathbf{Y}_m^*$.

Step 2: Using the penalized regression spline method, obtain a smoothing spline estimate $\hat{\mu}_m(t)$ based on the $\mathbf{Y}_m^*$ and then obtain the slope parameter estimate $\beta_{m,1}^*$ by fitting a linear model to $\hat{\mu}_m(t)$, i.e.

$$\hat{\mu}_m(t) = \beta_{m,0} + \beta_{m,1}^* t$$

Step 3: Repeat Steps 1 and 2 for M times, and the imputation-based estimator of the slope is derived as follows

$$\beta_1^* = \frac{1}{M}\sum_{m=1}^{M}\beta_{m,1}^*$$

The variance estimator of $\beta_1^*$ can be obtained as follows

$$\frac{1}{M}\sum_{m=1}^{M}W_m^{\beta} + \left(1+\frac{1}{M}\right)B_{\beta}$$

where $B_{\beta}$ is the between-imputation sample variance of $\beta_{m,1}^*$ (m = 1,…,M), and $W_m^{\beta}$ is the variance of $\beta_{m,1}^*$ within its specific imputed dataset which can be estimated by the block bootstrap approach where bootstrap samples were generated based on block bootstrap resamples of residuals from the fitted linear model.

To test whether there is significant M change in the slope from pre- to post-transfusion period, we first apply Algorithm 3 to estimate the pre- and post-transfusion slope parameters $\beta_{1,pre}^*$ and $\beta_{1,post}^*$ based on $\mathbf{Y}_{\text{pre}}$ and $\mathbf{Y}_{\text{post}}$, respectively. We then evaluate the pre- to post-transfusion changes in the slope as

$$\Delta_{slope} = \beta_{1,post}^* - \beta_{1,pre}^*$$

To derive the *p*-value for testing the significance of the change in slope, we apply an MI-nested permutation test procedure similar as the testing procedure presented in Algorithm 2. One slight difference with the slope test is that in the permutation procedure we permute the time points in pre-transfusion data $\mathbf{Y}_{\text{pre}}$ and the time points in the post-transfusion data $\mathbf{Y}_{\text{post}}$ separately to generate each permuted sample. This permutation strategy will help avoid potential confounding effects due to the difference in the mean $rSO_2$ levels in the pre- and post-transfusion periods when we test on the slope difference.

## 4 Simulation studies

We conducted simulation studies to evaluate the performance of the proposed methods in comparison with alternative methods. First, we performed a simulation study to evaluate the accuarcy of the proposed MAUC method in estimating the mean $rSO_2$ level between the pre- and post-transfusion periods. We simulated mesenteric $rSO_2$ series data that mimic the characteristics of the real $rSO_2$ series observed in the study. Specifically, we first generated the underlying smooth function $\mu(t)$ based on the biological process estimated from real $rSO_2$ time series, hence capturing realistic oxygenation temporal dynamics. We then generated the residual time series, which incorporate the random variations in the observed $rSO_2$ measurements around the underlying smooth function. Specifically, we adopted the linear process bootstrap (LPB) approach[22] to simulate residual time series by bootstrapping the real residual time series estimated from the observed $rSO_2$ data. As a technique developed for bootstrapping-dependent time series data, LPB has an important advantage in that it can generate residual time series that maintain the covariance structure in the real $rSO_2$ series. The simulated residual time series were then added to the smooth function to generate the observed NIRS $rSO_2$ measurements. As in the real data, we incorporated the detection limit by replacing any $rSO_2$ measurement below 15% with 15%. We considered three simulation settings with different combinations of pre-transfusion and post-transfusion mean $rSO_2$ levels. The percentage of observations at the detection limit ranged between 10% and 50% across the settings, which were in compliance with our observation from the real data. We applied the proposed MAUC method to estimate the mean $rSO_2$ level at pre- and post-transfusion periods and then estimate the difference between the two periods. In comparison, we considered the standard estimates based on sample means of the observed $rSO_2$ data. Furthermore, we evaluated the performance of the proposed MI-nested permutation test vs. the standard *t*-test for detecting changes in mean $rSO_2$ from pre- to post-transfusion. We compared both the type I error rate as well as the statistical power of the two tests.

**Table 1.** Simulation studies for estimating the mean rSO$_2$ level at pre- and post-transfusion period based on the sample mean method and the proposed MAUC method.

| | True value | Detection limit percentage | MAUC method BIAS (SD) | Sample mean method BIAS (SD) |
|---|---|---|---|---|
| Simulation Case 1 | | | | |
| Pre-trans. Mean rSO$_2$ | 17.709 | 30% | 0.017 (0.213) | 2.553 (0.312) |
| Post-trans. Mean rSO$_2$ | 17.709 | 30% | −0.016 (0.242) | 2.496 (0.371) |
| Difference (post−pre) | 0 | | −0.032 (0.302) | −0.057 (0.487) |
| Simulation Case 2 | | | | |
| Pre-trans. Mean rSO$_2$ | 17.709 | 30% | −0.031 (0.217) | 2.483 (0.336) |
| Post-trans. Mean rSO$_2$ | 20.724 | 10% | −0.034 (0.158) | 1.233 (0.291) |
| Difference (post−pre) | 3.015 | | −0.003 (0.270) | −1.250 (0.452) |
| Simulation Case 3 | | | | |
| Pre-trans. Mean rSO$_2$ | 15.062 | 50% | −0.003 (0.258) | 3.742 (0.393) |
| Post-trans. Mean rSO$_2$ | 20.724 | 10% | −0.051 (0.170) | 1.208 (0.266) |
| Difference (post−pre) | 5.662 | | −0.048 (0.310) | −2.535 (0.452) |

In the second simulation study, we evaluated the performance of the proposed slope measure. We first generated the underlying smooth function based on estimates from real rSO$_2$ time series. Specifically, we first generated a zero-slope $\mu_0(t)$ by detrending and demeaning an estimated smooth curve from real rSO$_2$ time series. We then added linear trends associated with specified slopes to $\mu_0(t)$ to generate $\mu(t)$ with temporal trajectories. Then, residual time series simulated using the LPB approach were added to $\mu(t)$ to generate the observed NIRS rSO$_2$ series. We considered three simulation settings with different combinations of pre-transfusion and post-transfusion slopes including: a no slope difference setting with no linear trend at either pre- and post-transfusion periods, a moderate slope difference setting with no linear trend at pre-transfusion and a positive linear trend after transfusion, and a large slope difference setting with a negative linear trend in pre-transfusion period and a positive trend in post-transfusion. We compared the accuracy of the proposed slope estimate in Algorithm 3 and the naïve slope estimate based on linear regression modeling of the rSO$_2$ observations. Furthermore, we evaluated the performance of the proposed slope test versus the naïve slope test in detecting pre- to post-transfusion changes in the slope.

Table 1 presents the results for estimating the mean rSO$_2$ levels between pre- and post-transfusion periods. Results show that as compared with the standard sample-mean based method, the proposed MAUC-based method demonstrated much lower bias in estimating the mean rSO$_2$ levels during the pre- and post-transfusion periods across all simulation scenarios. The proposed method also showed higher accuracy in estimating the pre- to post-transfusion changes in mean rSO$_2$ levels. Especially, when the percentage of observations at the detection limit was inbalanced between the pre- and post-transfusion periods, the standard method had a large bias in estimating the mean changes in rSO$_2$ while the proposed method demonstrated very small bias. Figure 1 presents the type I error and power in detecting the changes in mean rSO$_2$ levels based on the proposed MI-nested permulation test for MAUC and the standard t-test based on the differences in the sample means. Results show that the type I error of the proposed test was closer to the nominal level under the null when there was no difference between pre- and post-tranfusion mean rSO$_2$ level. When the pre- to post-transfusion difference in the mean rSO$_2$ increased, the power to detect the difference rised in both tests but the proposed test in general demonstrated much higher power as compared with the standard *t*-test.

Table 2 presents the results for estimating the slope of the rSO$_2$ series during pre- and post-transfusion periods. Results show that as compared with the naïve slope estimate, the proposed slope estimate demonstrated much lower bias in estimating the slope of the rSO$_2$ series during the pre- and post-transfusion periods. The proposed method also showed higher acuracy in estimating the changes in the slope from pre- to post-transfusion. Furthermore, for testing the changes in slope, Figure 2 shows the proposed testing procedure demonstrated type I error that was closer to the nominal level under the null when there was no pre- to post- transfusion difference. When there was a difference between the pre- and post-transfusion slopes, the proposed test procedure had better power in detecting the change in slope as compared with the naïve slope test.

Finally, we evaluate the performance of the proposed variance estimators for the MAUC and slope measures in several simulation settings (Table 3). Results show that the average of the proposed standard error

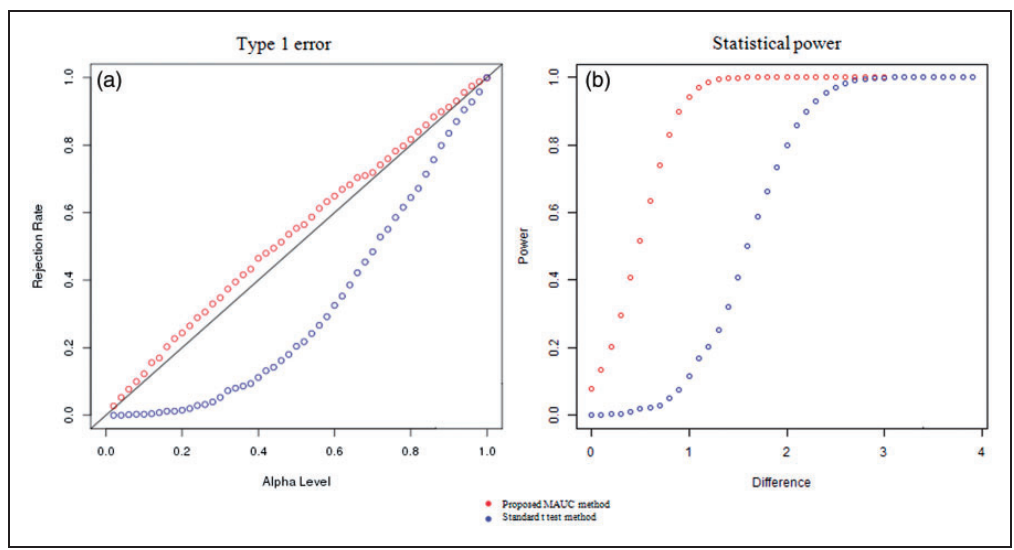

**Figure 1.** Simulation results for testing the changes in mean $rSO_2$ levels between pre- and post-transfusion using the proposed test for MAUC and the standard t-test. (a) We compared type I error rate at various alpha levels when there was no difference in the mean $rSO_2$ level between pre- and post-transfusion periods. The proposed test had a type I error that was closer to the nominal level. (b) We compared the power of the tests for detecting the differences in mean $rSO_2$ levels when alpha = 0.05. The proposed test demonstrated much higher power as compared with the standard t-test.

**Table 2.** Simulation studies for estimating the slope of $rSO_2$ series at pre- and post transfusion based on the naïve slope estimate and the proposed slope estimate.

|  | True values | The proposed slope method BIAS (SE) | The naïve slope method BIAS (SE) |
|---|---|---|---|
| Simulation Case 1 |  |  |  |
| Pre-trans. Slope $rSO_2$ | 0 | −0.08 (6.03) | −0.82 (8.7) |
| Post-trans. Slope $rSO_2$ | 0 | 0.24 (6.74) | −1.14 (10.22) |
| Difference (post–pre) | 0 | 0.16 (9.45) | −0.32 (12.52) |
| Simulation Case 2 |  |  |  |
| Pre-trans. Slope $rSO_2$ | 0 | −0.55 (6.51) | −1.48 (9.54) |
| Post-trans. Slope $rSO_2$ | 5 | −0.66 (5.26) | −4.88 (8.69) |
| Difference (post–pre) | 5 | −0.12 (8.49) | −3.39 (12.82) |
| Simulation Case 3 |  |  |  |
| Pre-trans. Slope $rSO_2$ | −5 | 0.38 (5.62) | 2.71 (9.80) |
| Post-trans. Slope $rSO_2$ | 5 | −0.51 (6.49) | −1.64 (9.07) |
| Difference (post–pre) | 10 | −0.90 (8.61) | −4.35 (14.04) |

Note: The slope values, bias and SE are on the scale of $10^{-3}$.

estimates was close to the Monte Carlo standard deviations of the MAUC and slope estimates, indicating accurate variance estimation.

## 5 Results for the Emory NIRS study

We applied the proposed methods to investigate the pre- and post-transfusion mesenteric $rSO_2$ among infants using already collected, de-identified data, from a previously published observational study conducted at Emory University.[19] First, we estimated mean $rSO_2$ at pre- and post-transfusion periods using the proposed MAUC measure based on Algorithm 1. The standard error of the MAUC estimates was obtained using the proposed variance estimator. We tested the difference between the pre- and post-transfusion MAUC using the MI-nested

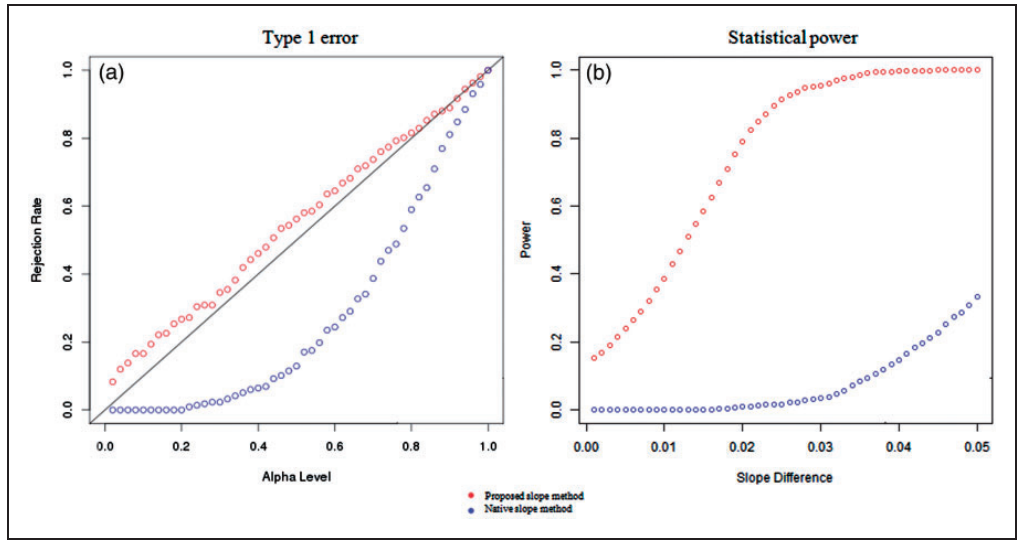

**Figure 2.** Simulation results for testing the changes in the slope between pre- and post-transfusion using the proposed slope test and the standard slope test based on linear regression modeling of the $rSO_2$ measurements. (a) We compared type I error rate at various alpha levels when there was no difference in the $rSO_2$ slope between pre- and post-transfusion periods. The proposed test had type I errors that were closer to the nominal levels. (b) We compared the power of the tests for detecting the differences in $rSO_2$ slopes between pre- and post-transfusion period. The proposed test demonstrated higher power as compared with the standard slope test.

**Table 3.** Simulation studies for standard error estimators for MAUC and Slope statistics.

| MAUC | MAUC SE estimator | | Slope | Slope SE estimator | |
|---|---|---|---|---|---|
| | Monto Carlo SD | The proposed SE | | Monto Carlo SD | The proposed SE |
| 17.709 | 0.511 | 0.509 (0.042) | 0 | 0.012 | 0.012 (0.001) |
| 19.166 | 0.508 | 0.534 (0.041) | 0.001 | 0.012 | 0.013 (0.001) |

Note: The slope values, its SE and SD are on the scale of $10^{-3}$.

permutation test in Algorithm 2. As a comparison with the MAUC method, we applied the existing method that evaluated the average oxygen level using the sample mean of the $rSO_2$ series and tested the pre-to-post-transfusion difference using the $t$-test. The standard error of the sample mean-based estimate for the average $rSO_2$ was obtained using the block bootstrap approach to take into account the temporal correlations in the $rSO_2$ series. Results in Table 4 show that the majority of the infants showed significant increase in the average oxygen level after the transfusion. When comparing the results from the standard sample mean method and the proposed MAUC method, we noted major differences in estimating the average $rSO_2$. Specifically, the sample mean tends to produce much higher estimate of the pre-transfusion $rSO_2$ as compared with the proposed MAUC method in some subjects. By examining the observed $rSO_2$ series of these subjects (Figure 3), we found they had a considerable amount of readings at the 15% detection limit prior to the transfusion. The sample mean method, which does not consider this limitation, overestimated the actual oxygen level in the subjects, while the proposed MAUC measure with the MI adjustment produced estimates that more accurately reflect the true tissue bed oxygenation. Since most of the infants' $rSO_2$ level increased following transfusion, the proportion of $rSO_2$ measurements occurring below the low detection limit generally decreased among the infants after transfusion. Therefore, the overestimation issue of the standard sample mean method tends to have a larger impact for an infant's pre-transfusion oxygen level than for his/her post-transfusion oxygen level. This in turn leads to biased estimates of the pre- to post-transfusion changes based on the standard method. That is, the standard sample mean method tends to underestimate the pre- to post-transfusion increases in oxygen level because it overestimates an infant's pre-transfusion level. The proposed MAUC method showed significant improvement in this issue. Across the infants who showed an increase in oxygen level after transfusion, the estimates of the pre- to post-transfusion changes in the rSO$_2$ level was 26.9% higher on average based on the proposed MAUC method as compared with the standard method. We note that the standard error of the sample mean estimates was sometimes lower than that of the proposed MAUC estimate especially for cases when there were a large amount of observations at the low detection limit. This is because with the standard sample mean methods, any rSO$_2$ measurements below 15% detection limit were cut off at 15%. Hence, its standard error estimates may underestimate the true variability of rSO$_2$.

**Table 4.** The average rSO$_2$ level in pre- to post-transfusion periods based on the sample mean method and the proposed MAUC method for infants from the Emory NIRS study.

| Subject[a] | The sample mean method | | | | The proposed MAUC method | | | |
|---|---|---|---|---|---|---|---|---|
| | Pre-tran. (SE) | Post-tran. (SE) | Pre- to post- change | _p_-Value[b] | Pre-tran. (SE) | Post-tran. (SE) | Pre- to post- change | _p_-Value[c] |
| A1 | 15.69 (0.64) | 52.44 (1.18) | 3.68 | <0.001 | 8.86 (0.93) | 52.18 (2.07) | 43.32 | <0.001 |
| A2 | 15.75 (0.63) | 27.66 (1.00) | 11.90 | <0.001 | 9.38 (0.86) | 25.46 (1.04) | 16.08 | <0.001 |
| B1 | 17.11 (0.76) | 16.04 (0.67) | −1.07 | 0.021 | 11.97 (1.17) | 8.98 (0.55) | −2.99 | <0.001 |
| B2 | 23.79 (1.06) | 42.13 (0.69) | 18.34 | <0.001 | 21.46 (1.60) | 41.02 (0.46) | 20.56 | <0.001 |
| C1 | 15.56 (0.56) | 17.08 (0.48) | 1.53 | <0.001 | 8.55 (0.75) | 11.65 (0.45) | 3.09 | <0.001 |
| C2 | 40.84 (1.01) | 49.08 (0.75) | 8.25 | <0.001 | 40.66 (1.19) | 49.11 (0.55) | 8.45 | <0.001 |
| D | 46.17 (1.03) | 35.32 (0.84) | −10.85 | <0.001 | 46.03 (1.63) | 34.63 (0.91) | −11.40 | <0.001 |
| E | 16.07 (0.59) | 22.99 (0.74) | 6.93 | <0.001 | 9.61 (1.26) | 20.22 (0.88) | 10.61 | <0.001 |
| F1 | 44.30 (0.86) | 44.16 (1.11) | −0.14 | 0.838 | 44.38 (1.06) | 43.20 (1.50) | −1.18 | 0.005 |
| F2 | 36.65 (1.05) | 49.38 (0.65) | 12.73 | <0.001 | 36.64 (0.88) | 49.40 (0.49) | 12.76 | <0.001 |
| G | 17.64 (0.77) | 19.80 (0.74) | 2.16 | <0.001 | 12.30 (1.07) | 15.43 (0.62) | 3.12 | <0.001 |
| H | 62.04 (1.85) | 50.58 (0.97) | −11.46 | <0.001 | 61.81 (3.27) | 50.45 (0.88) | −11.36 | <0.001 |
| I | 42.93 (0.94) | 51.55 (0.78) | 8.62 | <0.001 | 42.91 (0.93) | 51.47 (0.75) | 8.56 | <0.001 |
| J | 15.25 (0.39) | 29.84 (1.03) | 14.60 | <0.001 | 7.73 (0.55) | 27.87 (1.13) | 20.14 | <0.001 |
| K | 46.32 (1.65) | 56.25 (0.78) | 9.93 | <0.001 | 46.23 (2.61) | 55.96 (0.63) | 9.73 | <0.001 |
| L | 26.59 (1.09) | 32.30 (0.84) | 5.71 | <0.001 | 25.90 (1.36) | 31.77 (0.71) | 5.87 | <0.001 |
| M | 64.55 (1.37) | 18.92 (0.82) | −45.63 | <0.001 | 64.29 (2.06) | 13.82 (0.89) | −50.47 | <0.001 |
| N | 45.36 (0.93) | 43.76 (0.67) | −1.59 | <0.001 | 45.16 (0.70) | 43.64 (0.44) | −1.52 | <0.001 |

[a]Different letters in the Subject list corresponding to different infants. For an infant who received multiple transfusions, a numeric value is added after a subject's letter ID to denote different transfusions (e.g. A1 and A2 corresponding to NIRS data from the first and second transfusion from subject A).
[b]The _p_-value of the pre- to post-transfusion change was derived from the _t_-test.
[c]The _p_-value of the pre- to post-transfusion change was derived from the MI nested permutation test.

We also applied the proposed slope method to investigate the changes in the rSO$_2$ trajectory after transfusion. Using the method presented in Algorithm 3, we estimated the slope of rSO$_2$ trajectory for pre- and post-transfusion periods. The slope parameters and their standard errors are reported in Table 5. We then applied the MI-nested permutation test to examine the significance of the pre- to post-transfusion changes in the slope measure. In comparison with the proposed method, we applied the naïve approach that estimated the slopes by fitting linear models to the observed rSO$_2$ measurements directly. The standard error of the slope statistic from the naïve approach is obtained using a bootstrap method where the bootstrap samples were generated based on block bootstrap resamples of the residuals from the fitted linear models. From results in Table 5, we find that the two methods generally yielded consistent results in terms of the direction of the trajectory but the proposed method tends to produce more statistically significant results in detecting the changes in the slope. In particular, for some of the subjects, e.g. subject A1, there was a clear improvement in the trajectory in the rSO$_2$ after transfusion (Figure 3). However, the result based on the naïve slope test was not significant ($p = 0.269$). In comparison, our proposed test successfully detected a highly significant increase in the slope after the transfusion ($p < 0.001$). Another observation from the results is that the slope measure can provide helpful complementary information on rSO$_2$ patterns as compared with the MAUC measure. For example, some infants may experience a quick deterioration in the intestinal condition where the rSO$_2$ level which was originally in the normal range drops significantly. Since the decreases in rSO$_2$ happen in a very short time interval, the MAUC measure, which reflects the average rSO$_2$, may not be alarmingly low yet for these infants. The slope estimate, on the other hand, can help detect this fast decreasing trajectory in oxygenation and provide timely warning for physicians. Furthermore, by complementing the MAUC measure with the slope measure, we can obtain better understanding on how the rSO$_2$ changes after an RBC transfusion in preterm infants.

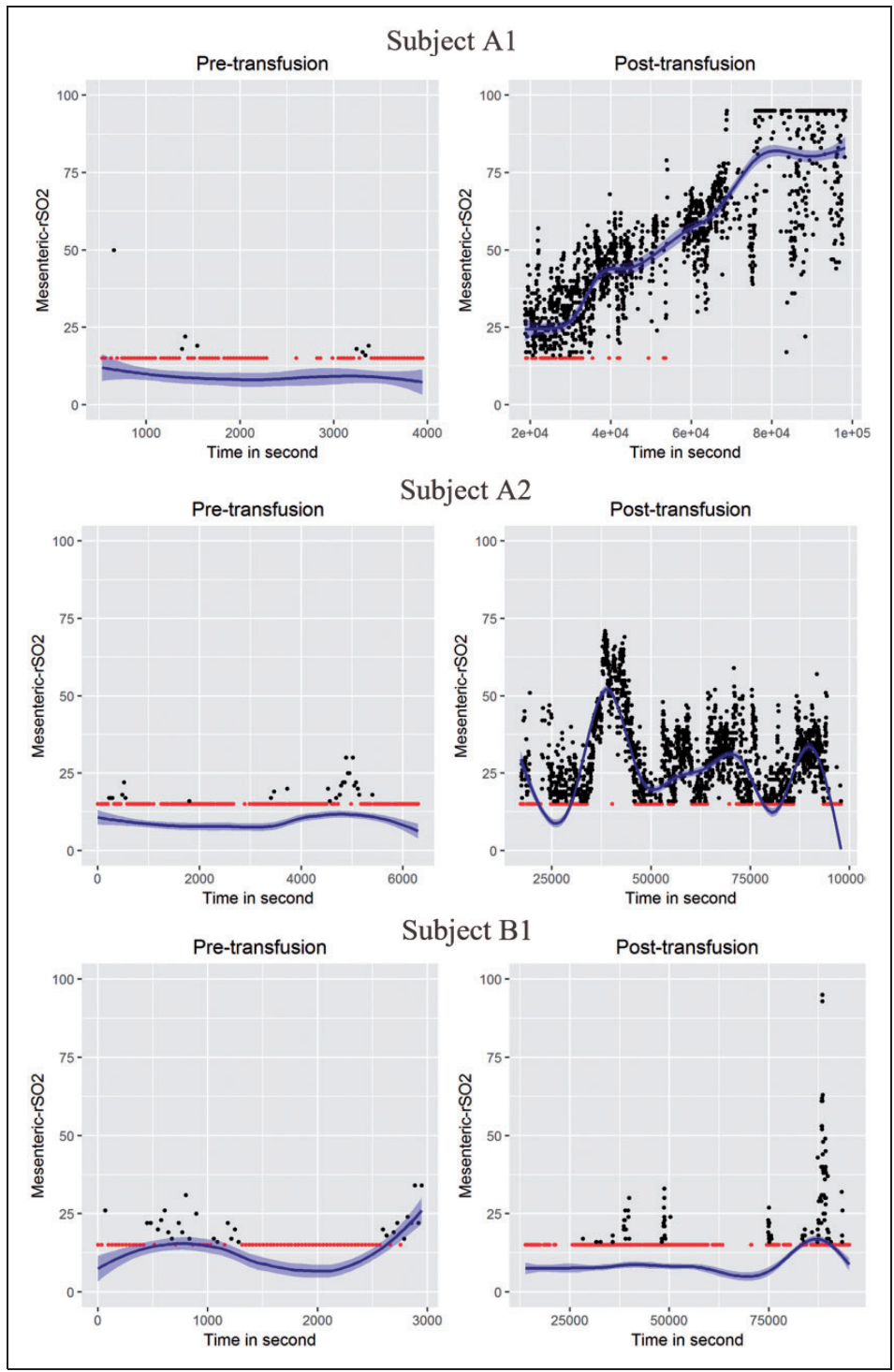

**Figure 3.** Observed pre- to post-transfusion rSO$_2$ series and the estimated underlying smooth functions for selected individual preterm infants from Emory NIRS study. The red dots are rSO$_2$ measurements occurring at the low detection limit of 15%.

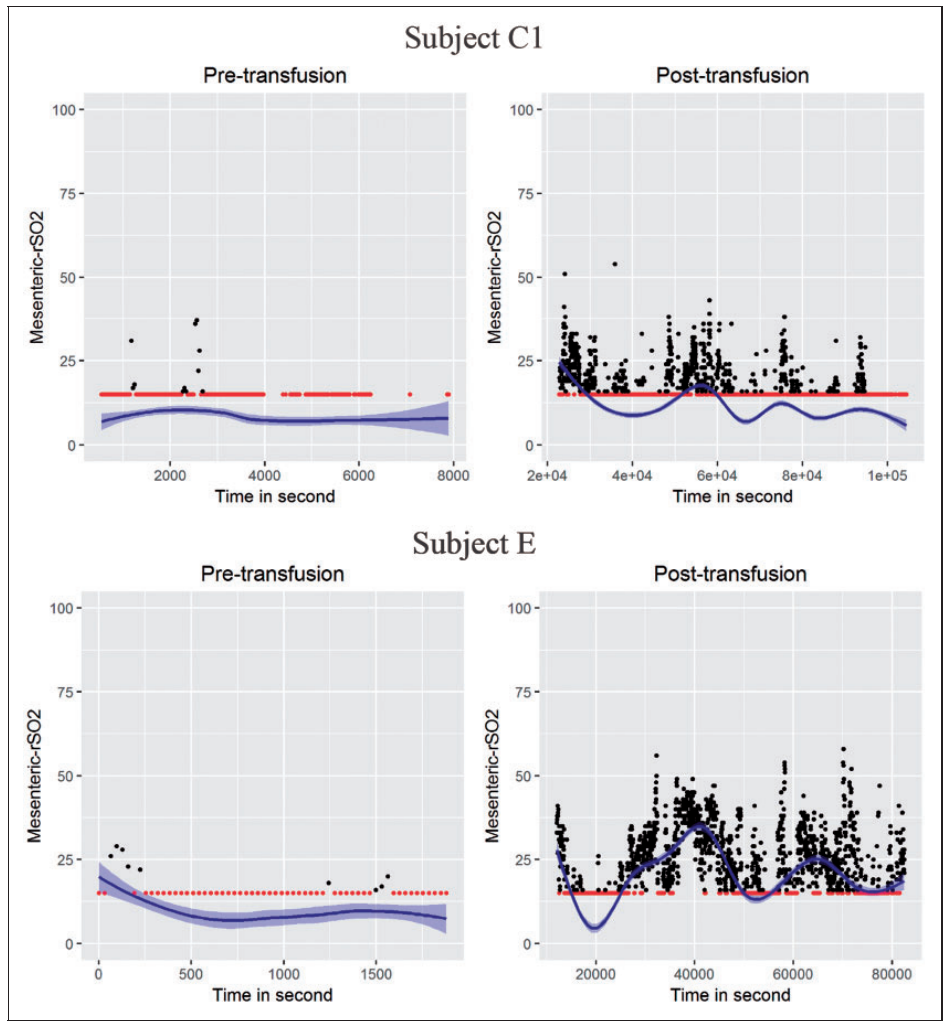

**Figure 3.** Continued.

**Table 5.** Pre- to post-transfusion changes in the rSO$_2$ temporal trajectory based on the naïve slope estimate and the proposed slope estimate.

| | The naïve slope method | | | | The proposed slope method | | | |
|---|---|---|---|---|---|---|---|---|
| Subject[a] | Pre-tran. (SE) | Post-tran. (SE) | Pre- to post- change | p-Value[b] | Pre-tran. (SE) | Post-tran. (SE) | Pre- to post-change | p-Value[c] |
| A1 | −0.549 (0.298) | 0.849 (0.035) | 1.398 | 0.269 | −0.419 (0.813) | 0.857 (0.035) | 1.276 | <0.001 |
| A2 | 0.210 (0.167) | −0.073 (0.040) | −0.283 | 0.570 | 0.333 (0.408) | −0.068 (0.045) | −0.401 | <0.001 |
| B1 | 0.357 (0.718) | 0.061 (0.019) | −0.296 | 0.641 | 0.656 (1.496) | 0.082 (0.023) | −0.574 | 0.684 |
| B2 | −0.923 (0.415) | −0.026 (0.020) | 0.897 | 0.000 | −1.486 (0.640) | −0.038 (0.020) | 1.448 | <0.001 |
| C1 | −0.190 (0.196) | −0.051 (0.013) | 0.139 | 0.466 | −0.447 (0.400) | −0.106 (0.019) | 0.341 | 0.004 |
| C2 | 0.377 (0.397) | −0.002 (0.018) | −0.379 | 0.038 | 0.349 (0.419) | −0.017 (0.023) | −0.366 | <0.001 |
| D | −3.207 (0.278) | −0.166 (0.039) | 3.041 | 0.000 | −3.129 (0.272) | −0.196 (0.034) | 2.933 | <0.001 |
| E | −2.054 (0.853) | 0.025 (0.028) | 2.079 | 0.317 | −2.930 (2.475) | 0.020 (0.040) | 2.951 | <0.001 |
| F1 | 0.731 (0.327) | −0.130 (0.072) | −0.861 | 0.012 | 0.823 (0.403) | −0.142 (0.069) | −0.965 | <0.001 |
| F2 | −0.101 (0.593) | 0.057 (0.022) | 0.158 | 0.736 | −0.051 (0.621) | 0.055 (0.022) | 0.106 | 0.622 |

(continued)

**Table 5.** Continued.

| | The naïve slope method | | | | The proposed slope method | | | |
|---|---|---|---|---|---|---|---|---|
| Subject[a] | Pre-tran. (SE) | Post-tran. (SE) | Pre- to post- change | $p$-Value[b] | Pre-tran. (SE) | Post-tran. (SE) | Pre- to post-change | $p$-Value[c] |
| G | 0.656 (0.312) | −0.076 (0.019) | −0.732 | 0.000 | 0.882 (0.448) | −0.111 (0.028) | −0.992 | <0.001 |
| H | −3.848 (0.567) | 0.030 (0.035) | 3.878 | 0.000 | −3.717 (0.660) | 0.015 (0.036) | 3.732 | <0.001 |
| I | 0.631 (0.238) | −0.104 (0.030) | −0.735 | 0.000 | 0.671 (0.262) | −0.097 (0.030) | −0.768 | <0.001 |
| J | −0.100 (0.104) | −0.136 (0.035) | −0.036 | 0.946 | −0.141 (0.272) | −0.165 (0.045) | −0.024 | 0.218 |
| K | −33.429 (4.16) | 0.037 (0.032) | 33.466 | 0.000 | −33.426 (3.99) | 0.029 (0.024) | 33.454 | <0.001 |
| L | −3.267 (0.270) | 0.153 (0.019) | 3.420 | 0.000 | −3.663 (0.320) | 0.146 (0.028) | 3.810 | <0.001 |
| M | −8.247 (4.272) | 0.158 (0.022) | 8.405 | 0.011 | −7.790 (4.381) | 0.240 (0.030) | 8.030 | <0.001 |
| N | 1.933 (0.430) | −0.024 (0.024) | −1.957 | 0.000 | 1.843 (0.395) | −0.033 (0.025) | −1.876 | <0.001 |

Note: The scale of the slope estimate and its SE is on $10^{-3}$.

[a]Different letters in the Subject list corresponding to different infants. For an infant who received multiple transfusions, a numeric value is added after a subject's letter ID to denote different transfusions.

[b]The $p$-value of the pre- to post-transfusion change was derived from $t$-test.

[c]The $p$-value of the pre- to post-transfusion change was derived from the MI nested permutation test.

# 6 Conclusions

In this study, we propose a new approach to analyse the mesenteric oxygenation response to red blood cell transfusion (RBC) in preterm infants. Our proposed approach addresses several limiations inherent in mesenteric NIRS monitoring including wide fluctuation in mesenteric signals, signal dropout and low limit of detection of the device. Compared with the standard methods currently applied to mesenteric NIRS measurements, the proposed methods demonstrate improved accuracy and increased power to detect RBC-related changes in mesenteric oxygenation in preterm infants. This approach has utility in assessing not only the response to RBC transfusion, but also other exposures such as feeding, anemia, and cardiovascular status on mesenteric oxygenation. Given the stability of NIRS montioring of other tissues beds such as cerebral oxygenation, our approach may not be as useful in the non-mesenteric setting. In addition, future prospective studies are needed to determine how the changes detected in rSO$_2$ relate to important clinical outcomes in preterm infants such as necrotizing enterocolitis.

## Declaration of conflicting interests

The author(s) declared no potential conflicts of interest with respect to the research, authorship, and/or publication of this article.

## Funding

The author(s) disclosed receipt of the following financial support for the research, authorship, and/or publication of this article: The study was funded by the NHLBI under grant no. P01 HL086773. The NIH had no role in the study design; the collection, analysis, and interpretation of data; the writing of the report; and (4) the decision to submit the paper for publication.

## ORCID iD

Ying Guo  http://orcid.org/0000-0003-3934-3097

## References

1. Sood BG, McLaughlin K and Cortez J. Near-infrared spectroscopy: applications in neonates. *Semin Fetal Neonatal Med* 2015; **20**: 164–172.
2. Marin T and Moore J. Understanding near-infrared spectroscopy. *Adv Neonatal Care* 2011; **11**: 382–388.
3. Mintzer J, Parvez B, Chelala M, et al. Quiescent variability of cerebral, renal, and splanchnic regional tissue oxygenation in very low birth weight neonates. *J Neonatal-Perinatal Med* 2014; **7**: 199–206.
4. Bailey SM and Mally PN. Review of splanchnic oximetry in clinical medicine. *J Biomed Optics* 2016; **21**: 091306.

5. da Costa CS, Greisen G and Austin T. Is near-infrared spectroscopy clinically useful in the preterm infant? *Archives of Disease in Childhood-Fetal and Neonatal Edition* 2015; 100: 558–561.
6. Lin PW and Stoll BJ. Necrotising enterocolitis. *Lancet* 2006; **368**: 1271–1283.
7. Neu J and Walker WA. Necrotizing enterocolitis. *New Engl J Med* 2011; **364**: 255–264.
8. Patel RM and Denning PW. Intestinal microbiota and its relationship with necrotizing enterocolitis. *Pediatric Res* 2015; **78**: 232.
9. Patel RM, Knezevic A, Shenvi N, et al. Association of red blood cell transfusion, anemia, and necrotizing enterocolitis in very low-birth-weight infants. *JAMA* 2016; **315**: 889–897.
10. Fortune P-M, Wagstaff M and Petros A. Cerebro-splanchnic oxygenation ratio (CSOR) using near infrared spectroscopy may be able to predict splanchnic ischaemia in neonates. *Intens Care Med* 2001; **27**: 1401–1407.
11. Gay AN, Lazar DA, Stoll B, et al. Near-infrared spectroscopy measurement of abdominal tissue oxygenation is a useful indicator of intestinal blood flow and necrotizing enterocolitis in premature piglets. *J Pediatric Surg* 2011; **46**: 1034–1040.
12. Epelman M, Daneman A, Navarro OM, et al. Necrotizing enterocolitis: review of state-of-the-art imaging findings with pathologic correlation. *Radiographics* 2007; **27**: 285–305.
13. Mohamed A and Shah PS. Transfusion associated necrotizing enterocolitis: a meta-analysis of observational data. *Pediatrics* 2012; **129**: 529–540.
14. Marin T, Josephson CD, Kosmetatos N, et al. Feeding preterm infants during red blood cell transfusion is associated with a decline in postprandial mesenteric oxygenation. *J Pediatr* 2014; **165**: 464–471.e461.
15. McNeill S, Gatenby J, McElroy S, et al. Normal cerebral, renal and abdominal regional oxygen saturations using near-infrared spectroscopy in preterm infants. *J Perinatol* 2011; **31**: 51.
16. Hastie T and Tibshirani R. *Generalized additive models*. New York, NY: Wiley Online Library, 1990.
17. Marin T, Moore J, Kosmetatos N, et al. Red blood cell transfusion–related necrotizing enterocolitis in very-low-birthweight infants: a near-infrared spectroscopy investigation. *Transfusion* 2013; **53**: 2650–2658.
18. Ruppert D, Wand MP and Carroll RJ. *Semiparametric regression*. Cambridge series in statistical and probabilistic mathematics 12. Cambridge: Cambridge University Press. Mathematical Reviews (MathSciNet): MR1998720 2003.
19. Hutchinson MF and De Hoog F. Smoothing noisy data with spline functions. *Numerische Mathematik* 1985; **47**: 99–106.
20. Succop PA, Clark S, Chen M, et al. Imputation of data values that are less than a detection limit. *J Occup Environment Hygiene* 2004; **1**: 436–441.
21. Hopke PK, Liu C and Rubin DB. Multiple imputation for multivariate data with missing and below-threshold measurements: time-series concentrations of pollutants in the Arctic. *Biometrics* 2001; **57**: 22–33.
22. McMurry TL and Politis DN. Banded and tapered estimates for autocovariance matrices and the linear process bootstrap. *J Time Series Analysis* 2010; **31**: 471–482.